\documentstyle[a4,12pt,epsfig]{article}
\def\nn{\nonumber}
\def\beq{\begin{equation}}
\def\eeq{\end{equation}}
\def\bea{\begin{eqnarray}}  \def\eea{\end{eqnarray}}
\def\lsim{\raise0.3ex\hbox{$<$\kern-0.75em\raise-1.1ex\hbox{$\sim$}}}
\def\gsim{\raise0.3ex\hbox{$>$\kern-0.75em\raise-1.1ex\hbox{$\sim$}}}
\def\1{{\rm 1\mskip-4.5mu l} }
\newcommand{\noi}{\noindent}
\begin{document}

\begin{center}
\vspace*{1 truecm}
{\Large \bf  Centrality Dependence of Hadron Multiplicities} \par 
\vskip 3 truemm
{\Large \bf in Nuclear Collisions in the Dual Parton Model} \\[8mm]

{\bf A. Capella and D. Sousa}\par
{\it Laboratoire de Physique Th\'eorique}\footnote{Unit\'e Mixte de 
Recherche - CNRS -
UMR N$^{\circ}$ 8627} \\ {\it Universit\'e
de Paris-Sud, B\^atiment 210, F-91405 Orsay Cedex, France}\\
[5mm]
{\bf }\par
{\it }
\end{center}

\vskip 1 truecm
\begin{abstract}
We show that, even in purely soft processes, the hadronic 
multiplicity in nucleus-nucleus
interactions contains a term that scales with the number of binary 
collisions. In the absence of
shadowing corrections, this term dominates at mid rapidities and high 
energies. Shadowing
corrections are calculated as a function of impact parameter and the 
centrality dependence of
mid-rapidity multiplicities is determined. The multiplicity per 
participant increases with
centrality with a rate that increases between SPS and RHIC energies, 
in agreement with experiment.
\end{abstract}

\vskip 4 truecm

\noindent LPT Orsay 00-137 \par
\noindent December 2000

\newpage
\pagestyle{plain}
\section{Introduction}
\hspace*{\parindent} Hadron multiplicities per unit rapidity at SPS 
energy show an approximate
scaling with the number of participants. This property is known as 
``wounded nucleon model'' (WNM)
\cite{1r}. However, a precise determination of the ratio $dN^{ch}/d\eta (\eta^*
= 0)/n_{part}$, both at CERN-SPS \cite{2r} and at RHIC \cite{3r}, 
shows a steady increase
from peripheral to central collisions. This increase is larger at 
RHIC energies, where the
data show no saturation for the most central bins.\par

It is a widespread belief that ``soft'' processes lead to a scaling 
in $n_{part}$ while ``hard''
ones lead to a scaling with the number $n$ of binary collisions. 
While, this is the case for total
cross-sections, it is not so for single particle inclusive 
cross-sections. If one neglects the
effects of shadowing (i.e. nuclear effects in structure functions 
associated to triple Pomeron
interactions), there is a theorem known as Abramovski, Gribov, 
Kancheli (AGK) cancellation
\cite{4r}, based on general principles, according to which 
$d\sigma^{ch}/dy$ scales with $n$ at
mid-rapidities and asymptotic energies. The best way to understand 
this cancellation is to
illustrate it with a model that satisfies these general principles, 
namely the Glauber model. \par

Let us consider for simplicity $pA$ scattering. The cross-section for 
$n$ inelastic collisions of
the proton with $n$ nucleons of the target is given by the 
probabilistic expression

\beq
\label{1e}
\sigma_n^{pA}(b) = {A \choose n} \Big ( \sigma_{pp} \ T_A(b) \Big )^n 
\Big ( 1 - \sigma_{pp} \
T_A (b) \Big )^{A-n} \quad .
\eeq

\noi where $T_A(b)$ are nuclear profile functions normalized to 
unity. Using this equation it is
easy to see that

\beq
\label{2e}
\sigma_{in}^{pA} = \int d^2b \sum_{n=1}^A \sigma_n^{pA}(b) \propto 
A^{2/3} \quad .
\eeq

\noi However, for the non-diffractive inclusive cross-section one has

\beq
\label{3e}
{d\sigma^{pA} \over dy} \propto \int d^2b \sum_{n=1}^A n 
\sigma_n^{pA}(b) \propto A^1 \quad .
\eeq
 
\noi Thus, one obtains the behaviour $A^1$ typical of hard processes. \par

The derivation of (\ref{3e}) assumes that the inclusive cross-section 
for $n$ inelastic
collisions is $n$ times the corresponding one for a single collision 
(or, more generally, that
the hadronic plateau produced in an inelastic collision does not 
depend on the number of
inelastic collisions). Clearly, such a condition is only
true at mid-rapidities and for asymptotic
energies. A basic idea in the WNM is that a nucleon (here the 
projectile), when wounded once,
looses its ability to produce extra particles in further collisions. 
This may be valid at low
energies, when the projectile undergoes successive collisions with the 
nucleons of the target.
However, at high energies, the space-time development of the 
interaction implies that the $n$
collisions are ``parallel'', i.e. they originate from different 
constituents of the projectile
wave function and take place simultaneously. (Technically, it means 
that one is dealing with
non-planar diagrams). This is the so-called Glauber-Gribov model 
\cite{5r}. A more detailed
discussion can be found in \cite{6r}.\par

As discussed above, at finite energies the constraints of 
energy-momentum conservation lead to
a violation of the AGK cancellation. The dual parton model (DPM) 
\cite{7r} and the quark gluon
string model (QGSM) \cite{8r}, while obeying AGK cancellation at mid 
rapidities and
asymptotic energies, contain an ``educated guess'' on its violation 
at finite energies. These
models are based on the quark-gluon content of hadrons in the 
framework of the Glauber-Gribov
model, and on the large $N$ expansion of non-perturbative QCD. The 
charged multiplicities per
unit rapidity are given by \cite{7r,9r}

  \bea
  \label{4e}
&&{dN_{AA}^{ch} \over dy}(y,b) = n_A (b) \left [ N_{\mu}^{qq^{P}-q_v^T}(y) +
N_{\mu}^{q_v^P-qq^T}(y) + (2k - 2) N_{\mu}^{q_s-\bar{q}_s} \right ] + \nn \\
&& \Big ( n(b) - n_A(b)\Big ) \Big ( 2 k \ N_{\mu}^{q_s-\bar{q}_s}(y) 
\Big ) \quad .   \eea

\noi  Here $P$ and $T$ stand for the projectile and target nuclei,

\begin{equation}
\label{5e}
n(b) =  \sigma_{pp} \int d^2s \ A^2 \ T_A(s) \ T_A(b - s) / 
\sigma_{AA}(b) \equiv \sigma_{pp} \
A^2 \ T_{AA}(b)/\sigma_{AA}(b) \end{equation}

\noi is the average number of binary collisions and

\begin{equation}
\label{6e}
n_A(b) \equiv n_{part}(b)/2 = \int d^2s \ A\ T_A(s) \ \left [ 1 - 
\exp (- \sigma_{pp}   \
A\ T_A(b - s) \right ]/\sigma_{AA}(b)
\end{equation}

\noi the average number of participants of nucleus $A$. $k$ is the 
average number of inelastic
collisions in $pp$ and $\mu (b) = kn(b)/n_A(b)$ is the average total 
number of collisions suffered
by each nucleon. The first term in (\ref{4e}) is the plateau 
height in a $pp$
collision, resulting from the superposition of $2k$ strings, 
multiplied by $n_A$. Since in DPM
there are two strings per inelastic collision, the second term, 
consisting of strings stretched
between sea quarks and antiquarks, makes up a total number of strings 
equal to $2kn$. \par

The hadronic multiplicities of the strings in (\ref{4e}) are obtained 
from a convolution of
momentum distribution function and fragmentation functions
(eqs. (3.1) to (3.4) of \cite{7r}). 
The former are given \cite{7r} by a product of Regge propagators
times a delta function of energy conservation:

\begin{equation}
\label{7e}
\rho_{kn}(x_{1},x_{2kn};x_{2},x_{3},\dots,x_{2kn-1})=
c_{kn}x^{-1/2}_{1}x^{3/2}_{2kn}x^{-1}_{2}\dots x^{-1}_{2kn-1}
\delta (1-\sum^{2kn}_{i=1}x_{i})
\end{equation}

\noindent
Here $x_{1}$ and $x_{2kn}$ denote the $x$--values of the valence
quark and diquark, respectively, $x_{2}\dots x_{2kn-1}$ those
of the sea quarks and antiquarks and $c_{kn}$ is obtained by
normalizing $\rho_{kn}$ to unity. The momentum distribution
of a single constituent is obtained by integrating (\ref{7e}) over
the $x_{i}$'s of all others. In order to regularize the 
singularities at $x_{i}\sim 0$ all $x_{i}$ ($i \ne 2kn$) in eq. (\ref{7e})
are replaced by \cite{10r} $\bar x_{i} = 
(x_{i}^{2} + 4 \mu^{2}/s)^{1/2}$ with $\mu^{2}$ = 0.1 GeV$^{2}$.
This introduces a free parameter $\mu$ in the model. The diquarks and
quark fragmentation functions into charged particles were determined
from $e^{+}e^{-}$ and/or $pp$ data. They are given by \cite{10r}

\begin{eqnarray}
z D_{qq}(z) & = & 1.12 (1-z)^{3} \\ \nonumber
zD_{q_{v}} = z D_{q_{s}} (z) & = & {1.12 \over 1.35} \ 
{1.3 (1-z)^{2} + 0.05 \over 1-0.5 z}
\end{eqnarray}

\noindent
Finally, we have taken for the threshold of $qq-q$ strings 
$s_{0} = (m_{p}+m_{\pi}^{T})^{2}$ and for the one of
$q-\bar{q}$ strings $s_{0} = (2m_{\pi}^{T})^{2}$, with
$m_{\pi}^{T}$ = 0.33 GeV.
The value 
of $k$ can be determined in a
generalized eikonal model. Alternatively, we can fix it in such a way 
that the plateau in $pp$
agrees with the experimental value of $dN^{ch}/dy (y^*=0)$ for 
non-diffractive events. These two
determinations are consistent with each other \cite{7r,8r}. \par

Note that at asymptotic energies, when the plateau height of all 
strings coincides (i.e.
$N^{qq-q_v}(y = 0) \simeq N^{q-\bar{q}_s}(y^* = 0))$ one recovers the 
AGK cancellation,
namely

\bea
\label{9e}
&&{dN_{AA}^{ch} \over dy} (b, y^* = 0) \sim n(b) 2k \ 
N^{q_s-\bar{q}_s}(y^*=0) \nn \\
&& \sim {A^2\ T_{AA}(b) \over
\sigma_{AA}(b)} \ \sigma_{pp}^{ND} {dN^{ND}_{pp} \over dy}(y^* = 0) = 
{A^2 \ T_{AA}(b) \over
\sigma_{AA}(b)} \ {d\sigma_{pp}^{ND}  \over dy} (y^*=0) \quad .\eea

\noi Already at RHIC energies, where $N^{q_s-\bar{q}_s}(y^* \sim 0)$ 
is substantially
smaller than $N^{qq-q}(y^* \sim 0)$, the last term of (\ref{4e}) 
turns out to dominate. However,
energy conservation constraints produce a decrease of 
$N_{\mu}^{q_s-\bar{q}_s}(y^* = 0)$ with
increasing centrality and, thus, the scaling in the number of binary 
collisions is not reached.
Since the $pp$ rapidity plateau $dN_{pp}^{ND}/dy = 
(1/\sigma_{pp}^{ND}) d\sigma^{ND}_{pp}/dy$
refers to non-diffractive events, for consistency we have to use also 
the non-diffractive
cross-section $\sigma_{pp}^{ND}$ (rather than the inelastic one) in 
the determination of $n_A$ and
$n$, in order to recover eq. (\ref{9e}) 
asymptotically\footnote{Conventionally one uses instead
$\sigma_{inel}$. In this case one should also use the $pp$ 
multiplicity for inelastic events.
Since the latter is about 10 \% smaller than the non-diffractive one 
\cite{11r}, the two effects
tend to compensate with each other and only a small difference 
remains in the calculated
multiplicities.}. \par

We can now compute the centrality dependence of hadronic 
multiplicities. We present first the
results obtained in the absence of shadowing, at three different 
energies~: $\sqrt{s} = 17.3$,
130 and 200 GeV. The corresponding non-diffractive cross-sections are 
\cite{11r,12r} $\sigma_{ND}
= 26$, 33 and 34 mb, respectively. We take $k = 1.4$, 2.0 and 2.2, 
corresponding to
$dN_{pp}^{ND}/dy = 1.56$, 2.72 and 3.04 \cite{11r,13r}. The values of the 
charged multiplicities
$N_{\mu}^{qq-q}(y^* = 0)$ and $N_{\mu}^{q-\bar{q}}(y^* = 0)$ of the 
individual strings calculated
in DPM, are listed in Table 1. The results are shown in Fig.~1 (solid 
line) and in Figs.~2-3
(dashed lines). At SPS we obtain a mild increase of the multiplicity 
per participant
consistent\footnote{The opposite claim was made in \cite{14r} based 
on a model which is an
over-simplified version of DPM.} with the results of the WA98 
collaboration \cite{2r} (see
Fig.~1). This increase gets stronger with increasing energies 
(Figs.~2 and 3).\par

It should be noted that the absolute value of the multiplicity at SPS 
energies determined in DPM
has some uncertainty. First, the excess of $K^+$ and $p$ over $K^-$ 
and $\bar{p}$ is
not properly taken into account in the above calculation. Second, the 
value of the multiplicity
in the $q_s-\bar{q}_s$ strings is affected by the value of its 
threshold.
The
latter has much less effect at $\sqrt{s} = 130$ and 200 GeV. However, 
shadowing corrections, which
are negligibly small at SPS energies, become important at RHIC and 
have to be taken into account.
This introduces some uncertainty at RHIC energies (see below).\par

As emphasized in \cite{15r}, shadowing corrections in Gribov theory 
are universal, i.e. they
apply both to soft and hard processes. They are closely related to 
the size of diffractive
production and, thus, are controlled by triple Pomeron diagrams 
\cite{15r,16r}. In the recent
papers \cite{17r} it was shown, that, when unitarity corrections are 
consistently taken into
account, one can describe hard diffraction at HERA and soft one 
(photoproduction) with the same
value of the triple Pomeron coupling. This value agrees with the one 
in ref. \cite{16r} and will be
used in the following calculation. The reduction of the multiplicity 
resulting from shadowing
corrections is given by the ratio \cite{15r}

\beq
\label{10e}
R_{AB}(b) = {\int d^2s\ f_A(s) \ f_B(b - s) \over T_{AB}(b)}
\eeq

\noi where

\beq
\label{11e}
f_A(b) = {T_A(b) \over 1 + AF(s)\  T_A(b)} \quad .
\eeq

\noi Here the function F is given by the integral of the ratio of the 
triple Pomeron
cross-section $d^2\sigma^{PPP}/dy dt$ at $t = 0$ over the single 
Pomeron exchange cross-section
$\sigma_p(s)$~:

\beq
\label{12e}
\left . F(s) = 4 \pi \int_{y_{min}}^{y_{max}} dy {1 \over 
\sigma_P(s)} {d^2 \sigma^{PPP} \over dy \
dt} \right |_{t=0} = C  \left [ \exp (\Delta y_{max}) - \exp (\Delta y_{min})
\right ] \eeq

\noi with $y = \ell n (s/M^2)$, where $M^2$ is the squared mass of 
the diffractive system. For a
particle produced at $y_{cm} = 0$, $y_{max} = {1 \over 2} \ell n ({s \over m_N^2})$ and $y_{min}
= \ell n (R_A m_N/ \sqrt{3})$. Using the parameters 
of \cite{14r} \cite{16r} we get~: $C =
0.31$~fm$^2$. \par

The values of the shadowing corrections at each impact parameter 
$R(b)$ in $Au$-$Au$ collisions at
$\sqrt{s} = 130$~GeV are shown in Table 1. Our results including 
shadowing are shown by the upper
lines of the dark bands in Figs.~2 and 3. \par

It should be stressed that our calculations refer to $dN/dy$ while 
the first RHIC measurements
\cite{3r} \cite{18r} refer to $dN/d\eta$. The latter is, of course, 
smaller at mid rapidities. This
difference is negligibly small as SPS where the laboratory 
pseudo-rapidity variable is used.
However, at $\sqrt{s} = 130$ and 200 GeV, where $\eta_{cm}$ is used 
instead, their ratio is as
large as 1.3 \cite{19r}\footnote{The increase of $<p_T>$ with centrality is
predicted to be rather small at RHIC \cite{20r}. Thus, we have used 
the same 1.3 reduction
factor for all centralities.}. This value is substantially larger 
than the value of 1.1 quoted
in \cite{3r}.\par

Our results for the centrality dependence of $dN/d\eta$ per 
participant at $\sqrt{s} = 130$~GeV
are shown in Fig.~4 (upper line of the dark band) and compared with 
the PHENIX data. As we see, the
centrality dependence is quite well reproduced. However, the absolute 
values are about 15 \%
higher than the data. \par

It should be stressed that the values of $R$, eq. (\ref{10e}) are 
quite large (see Table 2). As
pointed out in \cite{13r} they have a rather large uncertainty at 
RHIC energies. An alternative
calculation of $R$ in ref. \cite{13r}, based on a formalism that 
reproduces the nuclear effects in
DIS on nuclei, led to values of shadowing about 15~\% larger than the 
ones obtained here. Clearly,
with these larger values of the shadowing corrections we would obtain 
a quantitative agreement with
the PHENIX data. \par

Multiplying the values of $R_{Au \ Au}(b)$ in Table 2 by a factor 
0.85 we obtain the lower lines
of the dark bands in Figs.~2-4. These bands can be regarded as an 
estimate of the uncertainties
on the values of the calculated multiplicities due to uncertainties 
in the shadowing
corrections. Note that these uncertainties affect mostly the absolute 
values of the
multiplicities -- while their centrality dependence is determined 
quite unambigously within our
model. \par

The multiplicity per participant at
SPS energies increases by 1.15 between $b = 10$~fm and $b = 0$. The
corresponding increase at $\sqrt{s} = 130$~GeV and 200 GeV is 1.30 
and 1.31, respectively. As a
consequence of this saturation, the rise of
the central plateau in $Au$-$Au$ collisions between these two 
energies is close to the one in $pp$
collisions -- an interesting prediction of our model.\par

We would like to compare our results with the ones obtained in other 
approaches.
Many Monte Carlos based on or inspired by DPM or QGSM do contain a 
term proportional to the
number of binary collisions. However, in other approaches 
\cite{21r,22r} 
such a term is associated
with minijets. Of course minijets are produced at high energies. They 
have been incorporated in DPM
and modify the $p_T$ dependence of the model \cite{23r}. However, 
they do not affect the
multiplicities, since they play the same role as $q_s$-$\bar{q}_s$ 
strings and the
total number of such strings is controlled by unitarity. A comparison with
\cite{22r} shows that, while in this approach the multiplicity is 
given by a linear combination
of $n(b)$ and $n_{part}(b)$, with coefficients independent of $b$, in 
our case these coefficients
decrease with increasing centrality. More important, in our approach 
these coefficients are
calculated while in \cite{22r} they are fitted to the data at each energy.\par

An interesting estimate of the centrality dependence of
charged multiplicities in high density QCD \cite{24r} has been 
presented in \cite{22r}.
Surprisingly the result at $\sqrt{s} = 130$~GeV is almost identical 
to the one based on the minijet picture obtained in \cite{22r}.
However, this result relys entirely on the logarithmic dependence
of the gluon structure function of the nucleon on the saturation scale
$Q_{s}^{2}$, used in \cite{22r}
(see last paper of ref. \cite{24r} for a discussion on this point).
It is also in\-te\-res\-ting to remark that the 
centrality dependence of the
multiplicity per participant in the minijet model \cite{22r} gets 
stronger with increasing energy
-- due to an increase of the minijet fraction. On the contrary, in 
high density QCD the effect is
the opposite one, namely, the (partonic) multiplicity per participant 
depends less and less on
centrality when energy increases\footnote{This is due to the fact 
that the geometrical factor $\rho_{part}(b)$ in
$Q_s^2$, which depends strongly on centrality, is independent of 
energy, while the factor $x
\ G(x)$, which depends on impact parameter very mildly, increases 
with $s$. Therefore $\ell n Q_s^2
\propto \ell n [x \ G(x)]$ at very high energies.}.  This is an
interesting prediction of the high density QCD saturation model. 
However, in the RHIC energy
range, from $\sqrt{s} = 130$ to 200~GeV, this effect is negligeably 
small and the centrality
dependence of the multiplicity per participant is the same at these 
two energies. Interestingly,
the same result is obtained in DPM when shadowing corrections are 
taken into account. \\

\noi {\Large \bf Acknowledgments} \\

It is a pleasure to thank Y. Dokshitzer, A. Kaidalov, A. Krzywicki, 
A. Mueller, C. Pajares, C.
Salgado and D. Schiff for interesting discussions. D.S. thanks
Fundaci\'on Barrie de la Maza for
financial support.

\newpage
\centerline{\Large \bf Figure Captions :}
\vspace{1cm}

\noi {\bf Figure 1.} The values of $dN^{ch}/dy/n_{part}$ 
versus $n_{part}$ for 
$PbPb$ collisions at $\sqrt{s}
= 17.3$~GeV in the range $- 0.5 < y_{cm} < 0.5$ computed from eqs. 
(\ref{4e}) to (\ref{6e}),
compared with the WA98 data \cite{2r} for $dN/d\eta/n_{part}$. The 
difference between $dN/d\eta$
and $dN/dy$ is very small since the laboratory pseudo-rapidity is 
used (see main text). \\

\noi {\bf Figure 2.} The values of $dN^{ch}/dy/(0.5 n_{part})$ for 
$Au$-$Au$ collisions at
$\sqrt{s} = 130$~GeV in the range $- 0.35 <y_{cm} < 0.35$ computed 
from eqs. (\ref{4e}) to
(\ref{6e}) (dashed line). The upper line in the dark band,
is obtained after shadowing 
corrections computed from eqs.
(\ref{10e})-(\ref{12e}). The lower one is obtained with a different 
determination of
the shadowing corrections (see main text).  \\

\noi {\bf Figure 3.} Same as Fig.~2 for $\sqrt{s} = 200$~GeV. \\

\noi {\bf Figure 4.} Same as Fig.~2 for $dN^{ch}/d\eta_{c.m.}$. The 
dashed line (results without
shadowing) has been omitted here. The PHENIX data \cite{3r} are also 
shown (black circles and
shaded area).

\vskip 2 truecm

\centerline{\Large \bf Table Caption :}
\vspace{1cm}

Values of $N_{\mu}^{qq-q}$ and $N_{\mu}^{q-\bar{q}}$ in eq. 
(\ref{4e}) at $\sqrt{s} = 130$~GeV
computed in DPM at different values of the impact parameter. The 
values correspond to charged
particles per unit rapidity in the range $- 0.35 < y_{cm} < 0.35$. 
The corresponding shadowing
corrections $R_{Au \ Au}(b)$ computed from eqs. (\ref{10e}-\ref{12e}) 
are given in the third
column.

\newpage
\centerline{\bf Fig. 1}
\begin{figure}[hbtp]
\begin{center}
\mbox{\epsfig{file=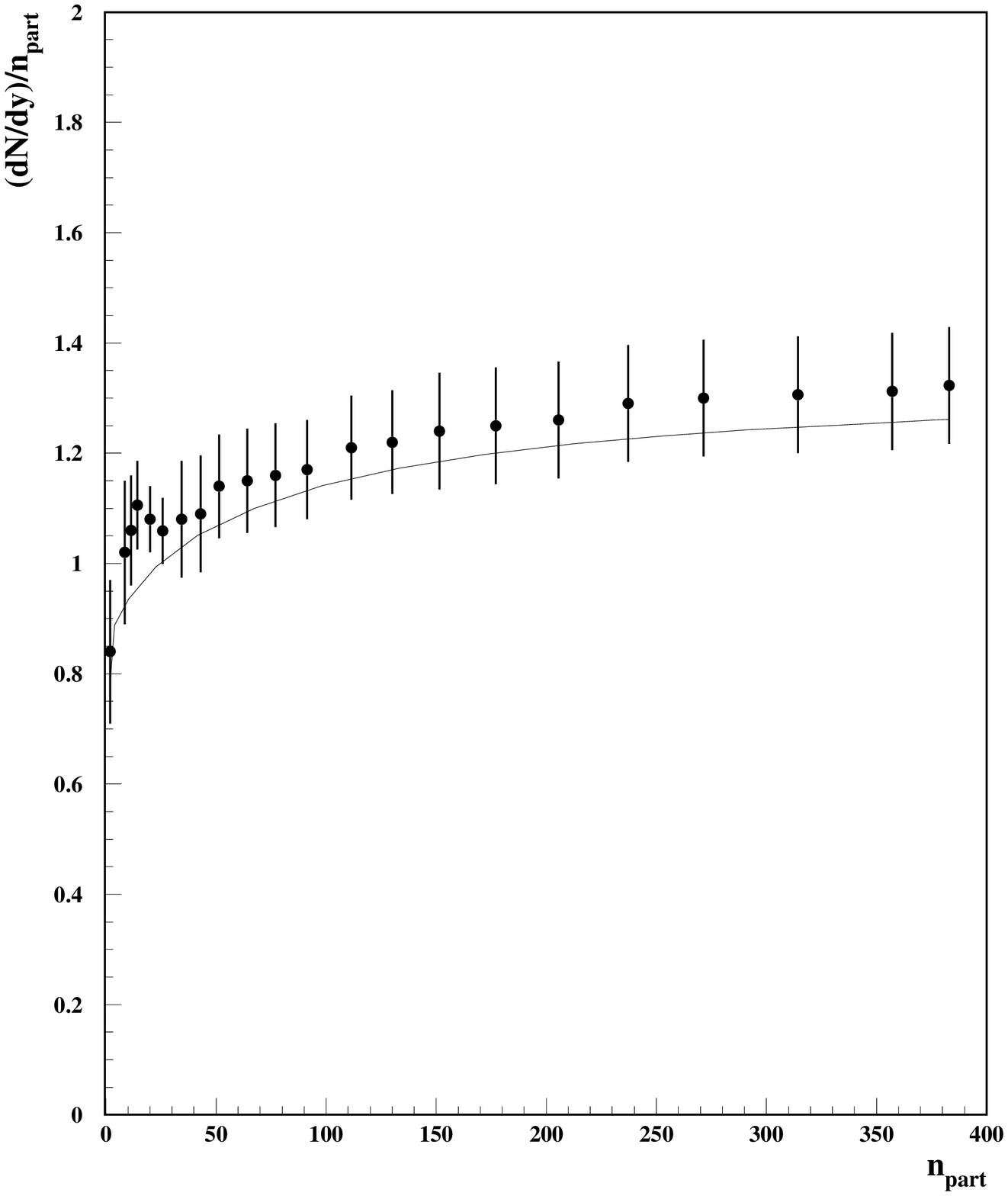,height=17cm}}
\end{center}
\end{figure}

\newpage
\centerline{\bf Fig. 2}
\begin{figure}[hbtp]
\begin{center}
\mbox{\epsfig{file=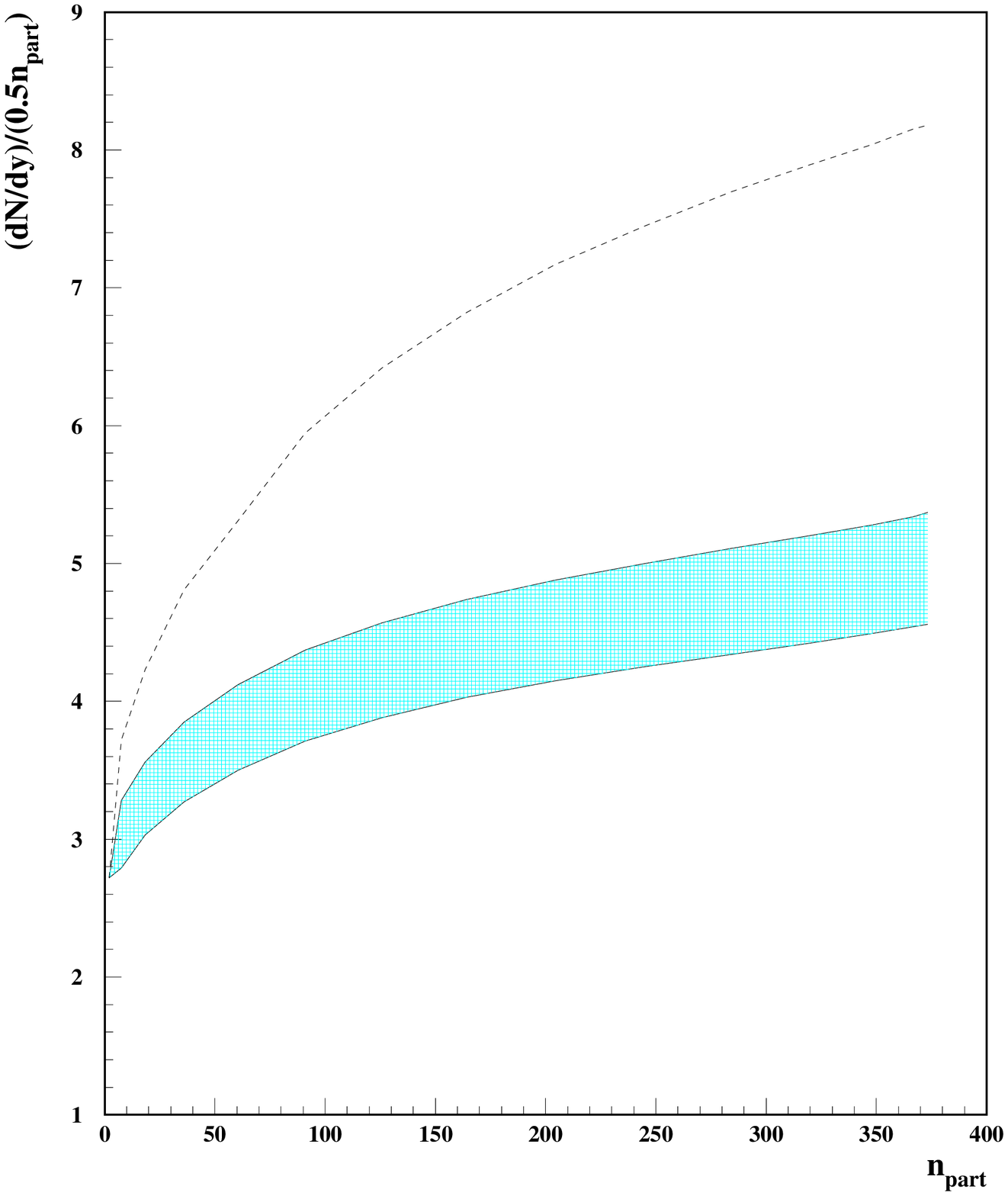,height=17cm}}
\end{center}
\end{figure}

\newpage
\centerline{\bf Fig. 3}
\begin{figure}[hbtp]
\begin{center}
\mbox{\epsfig{file=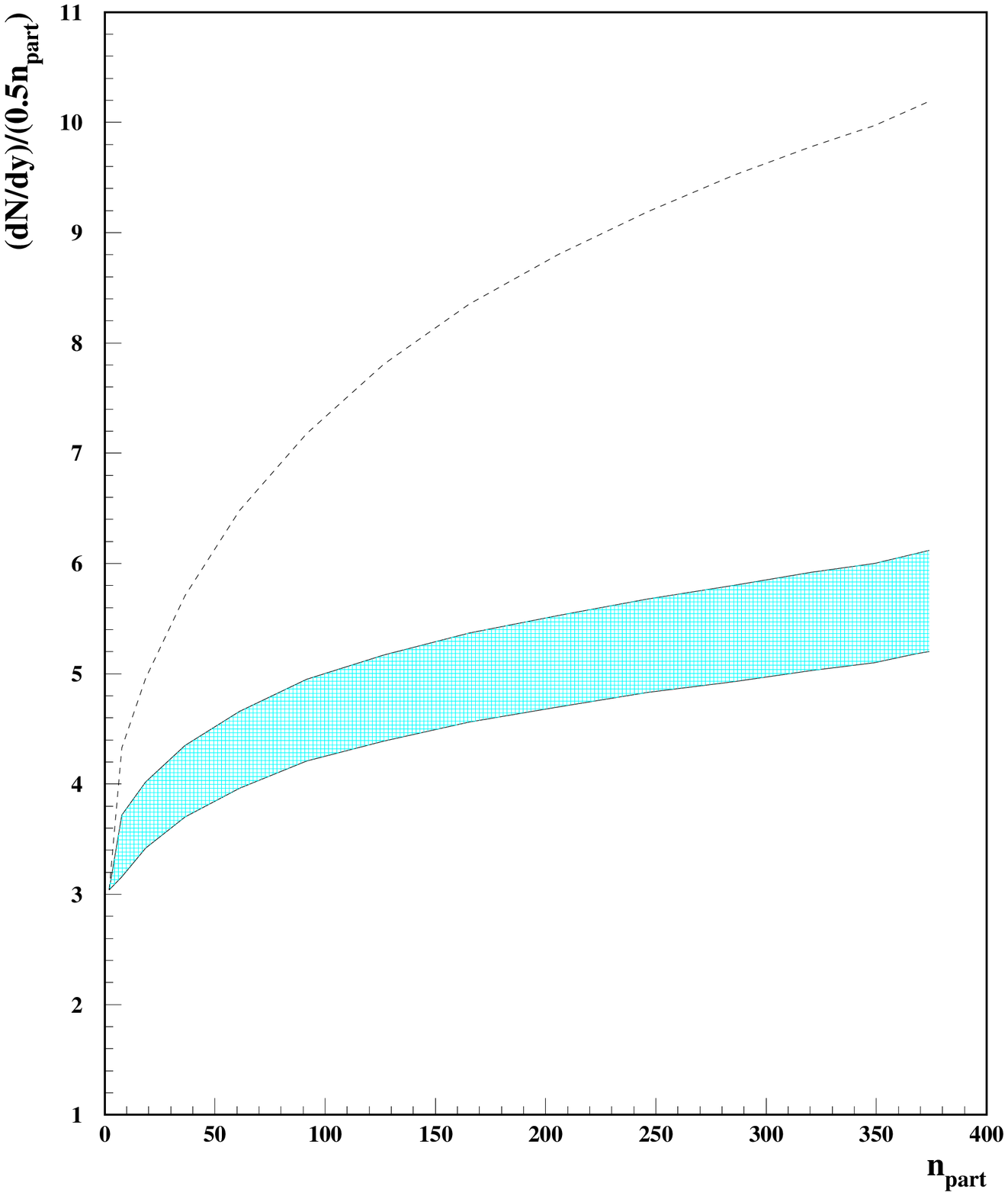,height=17cm}}
\end{center}
\end{figure}

\newpage
\centerline{\bf Fig. 4}
\begin{figure}[hbtp]
\begin{center}
\mbox{\epsfig{file=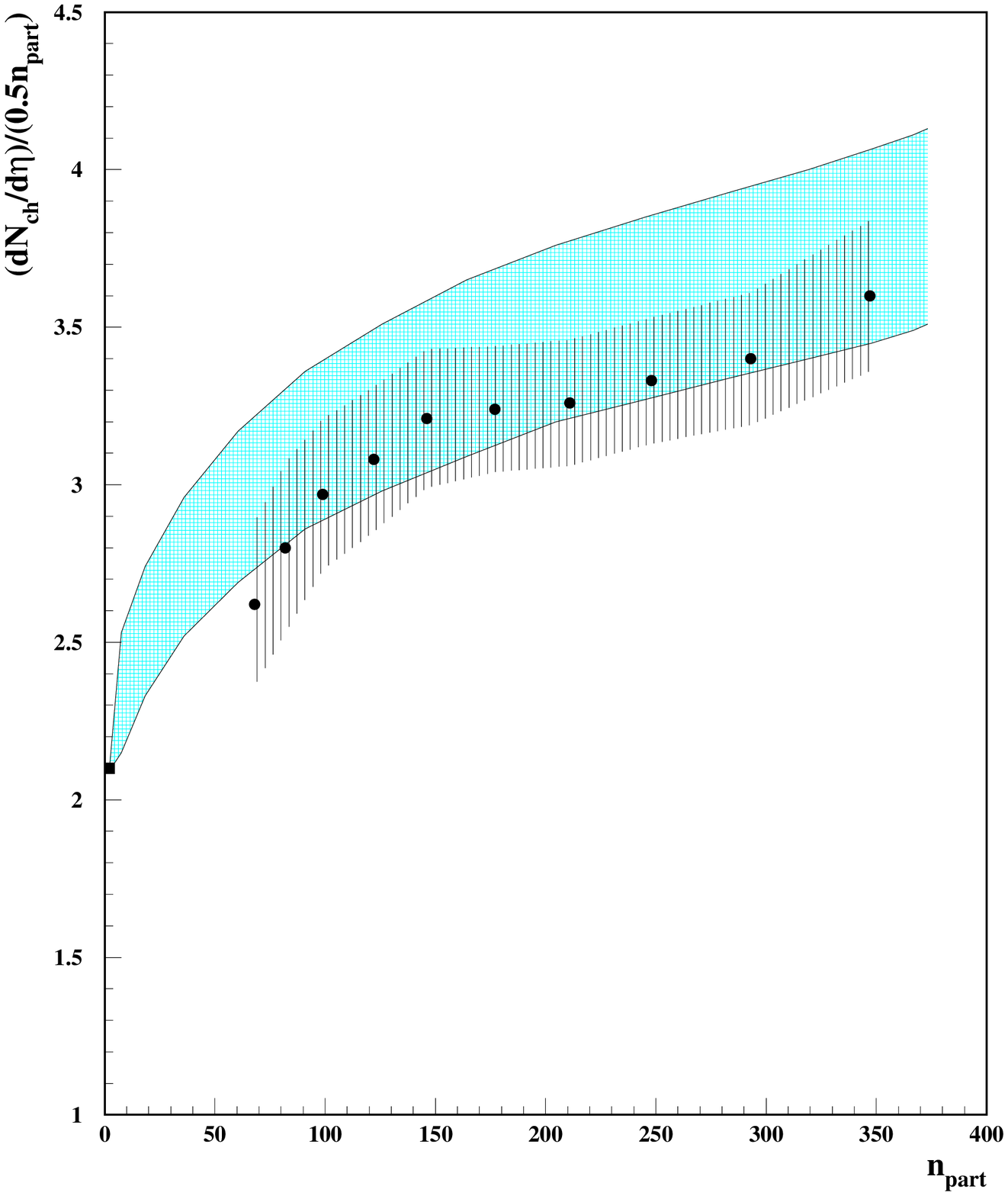,height=17cm}}
\end{center}
\end{figure}
 
\newpage
\begin{center}
\begin{tabular}{|c|c|c|c|}
\hline
& & & \\
\qquad $b$ (fm) \qquad &\qquad $N^{qq-q}$ \qquad &\qquad 
$N^{q-\bar{q}}$ \qquad &\qquad
$R_{Au-Au}$ \qquad \\
  & & & \\
\hline
& & & \\
0 &0.859 &0.345 &0.656 \\
2 &0.861 &0.347 &0.657 \\
4 &0.867 &0.351 &0.664 \\
6 &0.875 &0.358 &0.681 \\
8 &0.887 &0.368 &0.712 \\
10 &0.903 &0.381 &0.763 \\
12 &0.921 &0.397 &0.843 \\
& & & \\
\hline
\end{tabular}
\end{center}
\centerline{\bf Table 1}

\end{document}